\documentclass{amsart}
\usepackage{amssymb}
\usepackage{amsfonts}
\usepackage{graphicx}
\usepackage{epstopdf}
\usepackage{hyperref}

\begin{document}

	\begin{center}
		{\Large \bf Chern-Simons theory with the exceptional gauge group \\ as a \\
			refined topological string \\
			\vspace*{1 cm}
			
			{\large  R.L.Mkrtchyan\footnote{mrl55@list.ru}
			}
			\vspace*{0.2 cm}
			
			{\small\it Yerevan Physics Institute, 2 Alikhanian Br. Str., 0036 Yerevan, Armenia}
			
		}
		
	\end{center}\vspace{2cm}

{\small  {\bf Abstract.} We present the partition function of Chern-Simons theory with the exceptional gauge group on three-sphere  in the form of a partition function of the refined closed topological string with relation $2\tau=g_s(1-b) $ between single K\"ahler parameter $\tau$, string coupling constant $g_s$ and refinement parameter $b$, where $b=\frac{5}{3},\frac{5}{2},3,4,6$ for $G_2, F_4, E_6, E_7, E_8$, respectively. The non-zero BPS invariants $N^d_{J_L,J_R}$ ($d$ - degree) are $N^2_{0,\frac{1}{2}}=1, N^{11}_{0,1}=1$. Besides these terms, partition function of Chern-Simons theory contains term corresponding to the refined constant maps of string theory. 

Derivation is based on the universal (in Vogel's sense) form of a Chern-Simons partition function on three-sphere, restricted to exceptional line $Exc$ with Vogel's parameters  satisfying $\gamma=2(\alpha+\beta)$. This line contains points, corresponding to the all exceptional groups. The same results are obtained for $F$ line $\gamma=\alpha+\beta$ (containing $SU(4), SO(10)$ and $E_6$ groups), with the non-zero  $N^2_{0,\frac{1}{2}}=1, N^{7}_{0,1}=1$. 

In both cases refinement parameter $b$ ($=-\epsilon_2/\epsilon_1$ in terms of Nekrasov's parameters) is given in terms of universal  parameters, restricted to the line, by $b=-\beta/\alpha$.  }


\section{Introduction}

Partition function $Z$ of Chern-Simons theory with an arbitrary gauge group on 3d sphere was calculated exactly by Witten \cite{W1}. It appears to be the $S_{00}$ element of matrix $S$ of modular transformations of affine characters. In the papers \cite{MV1,M13} we have transformed this partition function into the universal form, i.e. expressed it in the terms of  Vogel's parameters $\alpha, \beta, \gamma$, which are homogeneous coordinates of  Vogel's plane (i.e. they are relevant up to rescaling and permutations) \cite{V0,V}. Their correspondence with the simple Lie algebras is given in Vogel's table \ref{tab:1}. 

\begin{table}[ht] 
	\caption{Vogel's parameters}
	\begin{tabular}{|r|r|r|r|r|r|} 
		\hline Algebra/Parameters & $\alpha$ &$\beta$  &$\gamma$  & $t$ & Line \\ 
		\hline  $SU(N)$  & -2 & 2 & $N$ & $N$ & $\alpha+\beta=0$ \\ 
		\hline $SO(N)$ & -2  & 4 & $N-4$ & $N-2$ & $ 2\alpha+\beta=0$ \\ 
		\hline  $ Sp(N)$ & -2  & 1 & $N/2+2$ & $N/2+1$ & $ \alpha +2\beta=0$ \\ 
		\hline $Exc(a)$ & $-2$ & $2a+4$  & $a+4$ & $3a+6$ & $\gamma=2(\alpha+\beta)$\\ 
		\hline 
	\end{tabular}
	
	{For the exceptional 
		line $Exc(a)$ $a=-2/3,0,1,2,4,8$ for $G_2, SO(8), F_4, E_6, E_7,E_8 $, 
		respectively.} \label{tab:1}
\end{table}

Universal partition function $Z$ is given by

\begin{eqnarray}\label{totalfree}
-\ln Z =
(dim/2)\ln(\delta/t)+
\int^{\infty}_0 \frac{dx}{x} \frac{f(x/\delta)-f(x/t)}{(e^{x}-1)} \label{Ftotal}
\end{eqnarray}

with

\begin{eqnarray}\label{gene}
f(x)&=&\frac{\sinh(x\frac{\alpha-2t}{4})}{\sinh(x\frac{\alpha}{4})}\frac{\sinh(x\frac{\beta-2t}{4})}{\sinh(x\frac{\beta}{4})}\frac{\sinh(x\frac{\gamma-2t}{4})}{\sinh(x\frac{\gamma}{4})} \\ \label{dim}
dim &=& \frac{(\alpha-2t)(\beta-2t)(\gamma-2t)}{\alpha\beta\gamma}\\ 
\delta&=&\kappa+t \\
t&=&\alpha+\beta+\gamma
\end{eqnarray}

Here $\kappa$ is a coupling constant of the Chern-Simons theory.  Scaling of Vogel's parameters change normalization of (otherwise unique) scalar product in simple Lie algebra, so  $\kappa$  should be scaled simultaneously to leave theory unchanged. At the so-called minimal normalization, when the length of long roots is equal to $2$ (parameters in table \ref{tab:1} are given in that normalization),  $\kappa$ becomes usual integer coupling constant $k$ of the Chern-Simons theory. In that normalization $t$ becomes dual Coxeter number, and $\delta=\kappa + t$ a usual shifted coupling constant.   $f(x)$ is the universal form of a quantum dimension of adjoint representation \cite{W3,MV1}, $dim$ is the dimension of simple Lie algebras in universal form. 

As it was recently shown \cite{KM}, (\ref{totalfree}) can be transformed into the even simpler form

\begin{eqnarray} \label{neat}
	-\ln Z=\frac{1}{4}\int_{R_+} \frac{dx}{x} \frac{\sinh\left(x(t-\delta)\right)}{\sinh\left( x t \right)\sinh\left(x \delta\right)} f(2x)
\end{eqnarray}

(\ref{totalfree}) is not merely the other form of the partition function. It actually insensibly analytically continues partition function to an arbitrary values of parameters (e.g. parameter $N$ for $SU(N)$ group). This is an ambiguous procedure for points from the Vogel table. However, it appears \cite{M13,KM} that exactly this analytical continuation is relevant for Chern-Simons/topological string duality. Particularly, the non-perturbative corrections to gauge \cite{M14SGV} and string \cite{HMMO13,H15} sides coincide, for $SU(N)$ case.

Further progress was achieved \cite{M13-2,M14SGV} by expressing  (\ref{totalfree}) for classical groups $SU, SO, Sp$ in terms of multiple sine functions (see also \cite{LV12, KM}). This form has an advantage that it can be transformed further into the sum over the residues of poles in the integral representation of the multiple sine functions (see \cite{Nar} for definition and properties of multiple sine functions). One series of these poles ("perturbative" ones) gives Gopakumar-Vafa expression for partition function of topological string, whereas the remaining poles give non-perturbative corrections.  For example, for $SU(N)$ gauge group one has: 

\begin{eqnarray} \label{SUN}
Z=\frac{\sqrt{d}}{ \sqrt{N}}  \frac{S_3(2N+2|2,2,2d) }{S_3(2|2,2,2d) }\\ \nonumber
d=k+N
\end{eqnarray}

Then one can consider the integral representation of the multiple sine functions:

\begin{eqnarray}\label{sinpol} \nonumber
S_r(z|\underline{\omega})=\\ 
\exp\left( (-1)^r \frac{\pi i}{r!}B_{rr}(z|\underline{\omega})+(-1)^r \int_{R+i0}\frac{dx}{x}\frac{e^{zx}}{\prod_{k=1}^r(e^{\omega_i x}-1)} \right)=\\ \nonumber
\exp\left((-1)^{r-1} \frac{\pi i}{r!}B_{rr}(z|\underline{\omega})+(-1)^r \int_{R-i0}\frac{dx}{x}\frac{e^{zx}}{\prod_{k=1}^r(e^{\omega_i x}-1)} \right)
\end{eqnarray}
where $B_{rr}$ are generalized Bernoulli polynomials \cite{Nar}. 

In the certain range of parameters' values it is possible to close the contour of integration in (\ref{sinpol}) in the upper semiplane and obtain an expression for (logarithm of) the sine function as the sum of the poles. 
As shown in \cite{M13-2} for the classical groups, multiple sine functions in representation of $Z$ in terms of sine functions (as (\ref{SUN})) have  one ray of poles (originated from multiplier $1/(e^x-1)$ in (\ref{totalfree})), which (ray) gives the corresponding  Gopakumar-Vafa form of the partition function.  For $SU(N)$ CS theory, i.e. for the partition function (\ref{SUN}) these "perturbative" poles come from parameter 2d multiplier in the integral representation (\ref{sinpol})  of triple sine. Contribution of  these poles is \cite{M13-2}

\begin{eqnarray} \label{SUNGV}
\ln Z \cong -\sum_{n=1}^{\infty}\frac{e^{-\tau n} }{n (e^\frac{g_s n}{2}-e^{-\frac{g_s n}{2}})^2}+\sum_{n=1}^{\infty}\frac{1}{n(e^\frac{g_s n}{2}-e^{-\frac{g_s n}{2}})^2}
\end{eqnarray}
where $g_s=2\pi i /d$, $\tau = -Ng_s$.

Here the first term is  the Gopakumar-Vafa  partition function of the topological string on resolved conifold \cite{GV,Mar1}, and the second term is the contribution of constant maps \cite{BCOV,GP,FP}.

Similar results are obtained for the refined $SU(N)$ and $SO(2n)$ cases: generalization of (\ref{totalfree})  was presented in  \cite{KS}, and poles expansion  was calculated in \cite{KM}. For the $SU(N)$ gauge group refined Gopakumar-Vafa plus constant maps part of (logarithm of) the partition function is \cite{KM}:

\begin{eqnarray} \label{FAPconifold}
 -\frac{1}{4}\sum_{n=1}^\infty \frac{Q^n}{n \sinh\left(\frac{n g_s}{2}\right)\sinh\left(\frac{n g_s b}{2}\right)}+\frac{1}{4}\sum_{n=1}^\infty\frac{ e^{- n g_s (b-1)/2}}{n \sinh\left(\frac{n g_s}{2}\right)\sinh\left(\frac{n  g_s b}{2}\right)}\,,
\end{eqnarray}

Note the deformation of numerator of constant maps term.

\section{Chern-Simons theories with exceptional groups} 

The aim of the present paper study is to extend this approach to the Chern-Simons theories with exceptional groups. We have  already commenced this work in \cite{M14SGV}, and will use now these results (rechecked for safety).

Approach to the exceptional groups is based on the fact, discovered by Vogel, that, as seen from table \ref{tab:1}, exceptional algebras belong to the line in Vogel's plane, as do linear or orthogonal/symplectic algebras. Based on this fact, Deligne \cite{Del96} suggested that this exceptional line has features, analogous to $SL$ or $SO/Sp$ lines on the same plane, i.e. in some definite sense all  points on that line have  similar features. In our approach we also treat them uniformly. 

We choose the parameterization of exceptional line 
$\alpha=z, \beta=1-z, \gamma=2$, with $z=-\frac{3}{5},-\frac{2}{5},-\frac{1}{3},-\frac{1}{4},-\frac{1}{6}$ for $G_2, F_4, E_6, E_7, E_8$, respectively, and perform similar calculations starting with the universal expression \ref{totalfree}.   According to \cite{M14SGV}, partition function on the exceptional line can be represented in terms of double sine function

\begin{eqnarray}
Z_{Exc}=   \frac{1}{4\pi\sin\frac{\pi}{2\delta}} \sqrt{\frac{-z}{y}}
\prod_p \left( \frac{ S_2(\frac{p}{2\delta}|1,\frac{y}{2\delta})}{S_2(\frac{p}{2\delta}|1,-\frac{z}{2\delta}) }\right)^{c_p} 
\end{eqnarray}
where $c_p$ is six-terms sequence $1,2,2,2,1,1$ for $p=1,2,3,4,5,6$, respectively.

One can represent this in terms of triple sine function, due to the relation 

\begin{eqnarray}
S_2(x|a_1,a_2)=\frac{S_3(x+a_3|a_,a_2,a_3)}{S_3(x|a_,a_2,a_3)}
\end{eqnarray}

\begin{eqnarray}
Z_{Exc}=   \frac{1}{4\pi\sin\frac{\pi}{2\delta}} \sqrt{\frac{-z}{y}}
\prod_p \left( \frac{ S_3(\frac{p}{2\delta}-\frac{z}{2\delta}|1,\frac{y}{2\delta},-\frac{z}{2\delta})}{S_3(\frac{p}{2\delta}+\frac{y}{2\delta}|1,\frac{y}{2\delta},-\frac{z}{2\delta}) }\right)^{c_p} 
\end{eqnarray}

Similar result holds for so-called F-line $\gamma=\alpha+\beta$. Let's parameterize this line as $\alpha=z, \beta=1-z, \gamma=1$. Points  $z=-\frac{1}{2},-\frac{1}{3},-\frac{1}{4}$ correspond to  $SU(4), SO(10)$, and  $E_6$, respectively. Partition function is similar to the $Exc$ line:

\begin{eqnarray}
Z_F=   \frac{\sqrt{\delta}}{2\sqrt{2\pi}\sin\frac{\pi}{2\delta}} \sqrt{\frac{-z}{y}}
\prod_p \left( \frac{ S_2(\frac{p}{2\delta}|1,\frac{y}{2\delta})}{S_2(\frac{p}{2\delta}|1,-\frac{z}{2\delta}) }\right)^{c_p} 
\end{eqnarray}
where now $c_p$ is sequence of four numbers $2,3,2,1$ for $p=1,2,3,4$, respectively. 

Similarly, in terms of triple sine partition function is

\begin{eqnarray}
Z_F =   \frac{\sqrt{\delta}}{2\sqrt{2\pi}\sin\frac{\pi}{2\delta}} \sqrt{\frac{-z}{y}}
\prod_p \left( \frac{ S_3(\frac{p}{2\delta}-\frac{z}{2\delta}|1,\frac{y}{2\delta},-\frac{z}{2\delta})}{S_3(\frac{p}{2\delta}+\frac{y}{2\delta}|1,\frac{y}{2\delta},-\frac{z}{2\delta}) }\right)^{c_p} 
\end{eqnarray}

Now, as in above formulae, for both $Exc$ and $F$ lines we take the integral representations of multiple sine functions, close contours in upper semiplane, and take contributions of the "perturbative" poles, only. Omitting elementary functions multiplier in the partition function, and Bernoulli polynomials from sine functions in the integral representation (\ref{sinpol}), we get the free energy

\begin{eqnarray} \label{a01}
\ln Z_{Exc,F} \cong \sum_p \sum_{n=1}^{\infty} \frac{c_p}{n} \left(  \frac{e^{2\pi i n \frac{p}{2\delta}}}{e^{2\pi in\frac{1-z}{2\delta}}-1}   - \frac{e^{2\pi i n \frac{p}{2\delta}}}{e^{2\pi in\frac{-z}{2\delta}}-1}  \right) = \\
\sum_p \sum_{n=1}^{\infty}\frac{c_p}{n} \frac{ e^{ n g_s ( -\frac{p}{z} +\frac{1}{2z} )} - e^{ n g_s ( -\frac{p}{z} -\frac{1}{2z} )  } }{ (e^{\frac{g_s n}{2}}-e^{-\frac{g_s n}{2}})  (e^{\frac{g_s b n}{2}}-e^{-\frac{g_s b n}{2}})}  
\end{eqnarray}

where we introduced the string coupling 

\begin{eqnarray}
g_s=-\frac{i\pi z}{\delta}
\end{eqnarray}

and refinement parameter

\begin{eqnarray}
b=1-\frac{1}{z}
\end{eqnarray}

Unrefined theory corresponds to 

\begin{eqnarray}
b\rightarrow +1, \,\, z \rightarrow - \infty
\end{eqnarray}

Using specific values of parameters $c_p$, we get for exceptional line $Exc$

\begin{eqnarray} \label{a01}
\ln Z_{Exc} \cong 
\sum_{n=1}^{\infty} \frac{e^{-\frac{g_s n}{2z}}+e^{-3\frac{g_s n}{2z}}-e^{-9\frac{g_s n}{2z}}-e^{-13\frac{g_s n}{2z}} } {n (e^{\frac{g_s n}{2}}-e^{-\frac{g_s n}{2}})  (e^{\frac{g_s b n}{2}}-e^{-\frac{g_s b n}{2}})}
\end{eqnarray}

and for F-line

\begin{eqnarray}
\ln Z_F \cong
\sum_{n=1}^{\infty} \frac{2e^{-\frac{g_s n}{2z}}+e^{-3\frac{g_s n}{2z}}-e^{-5\frac{g_s n}{2z}}-e^{-7\frac{g_s n}{2z}}-e^{-9\frac{g_s n}{2z}} } {n (e^{\frac{g_s n}{2}}-e^{-\frac{g_s n}{2}})  (e^{\frac{g_s b n}{2}}-e^{-\frac{g_s b n}{2}})} 
\end{eqnarray}

We would like to compare these partition functions with general form of the partition function of the refined topological string \cite{IKV07,N02}:
\begin{quotation} \tiny
\begin{eqnarray} \label{ztop}
&\ln Z_{top} \cong \\ \nonumber
&-\sum_{C\in H_2(X,\mathbb{Z})} \sum_{n=1}^{\infty}\sum_{j_L,j_R} \frac{(-1)^{2j_L+2j_R}N^C_{j_L,j_R}((qt)^{-nj_L}+...+(qt)^{nj_L})((\frac{q}{t})^{-nj_R}+...+(\frac{q}{t})^{nj_R})}{n(q^{\frac{n}{2}}-q^{-\frac{n}{2}})(t^{\frac{n}{2}}-t^{-\frac{n}{2}})} e^{-nT_C}
\end{eqnarray}
\end{quotation}
 
 Here $q=e^{\epsilon_1}, t=e^{-\epsilon_2}$, in terms of Nekrasov's parameters $\epsilon_1, \epsilon_2$. $T_C$ is the mass of brane wrapping cycle $C\in H_2(X,\mathbb{Z})$, with spin w.r.t. the $SU(2)_L \times SU(2)_R$ given by half-integers $j_L,j_R$. $N^C_{j_L,j_R}$ is the number of corresponding BPS multiplets, so it should be a positive integer number.  We put minus sign in front of the expression (\ref{ztop}) to have a coincidence with Chern-Simons theory SU(N) in an unrefined case at $q=t, N^1_{0,0}=1$. Finally, one also have to add to (\ref{ztop}) the term, originating from the constant maps.
 
 So, we want to present Chern-Simons partition functions in the form (\ref{ztop}) and  identify parameters of two theories. Let us rewrite the partition function for the $F$ line:

 \begin{eqnarray}
 \ln Z_F \cong \sum_{n=1}^{\infty} \frac{e^{-2\frac{g_s n}{2z}}(e^{-\frac{g_s n}{2z}}+e^{\frac{g_s n}{2z}})-e^{-7\frac{g_s n}{2z}}(e^{-2\frac{g_s n}{2z}}+1+e^{2\frac{g_s n}{2z}}) } {n (e^{\frac{g_s n}{2}}-e^{-\frac{g_s n}{2}})  (e^{\frac{g_s b n}{2}}-e^{-\frac{g_s b n}{2}})} \\ \label{cmtF}
 +\sum_{n=1}^{\infty} \frac{e^{-\frac{g_s n}{2z}}} {n (e^{\frac{g_s n}{2}}-e^{-\frac{g_s n}{2}})  (e^{\frac{g_s b n}{2}}-e^{-\frac{g_s b n}{2}})}
 \end{eqnarray}
 
 Now we shall identify parameters as shown below.  
 
\begin{eqnarray} 
q=e^{g_s},\,\,\,t=e^{g_s b} \\ \label{ttg}
T_C\equiv \tau = \frac{g_s(1-b) }{2} \\   \label{Flineident}
N^{2B}_{0,\frac{1}{2}}=1, \,\,\, N^{7B}_{0,1}=1
\end{eqnarray}

We assume that  the Calabi-Yau threefold has one K\"ahler parameter $\tau$, which is a value of K\"ahler form on basic cycle $B\in H_2(X,\mathbb{Z})$  and non-zero are numbers $N^C_{j_L,j_R}$ at $C=2B$ and $C=7B$, i.e. at the degrees 2 and 7.  Then two partition functions coincide, besides the constant maps term (\ref{cmtF}).
This last term coincides exactly with constant map term from refined $SU(N)$ theory \ref{FAPconifold}. From this we deduce that Euler characteristic of corresponding manifold $X$ is the same as in (\ref{SUNGV}), i.e. $\chi(X)=2$. 

Thus, for universal Chern-Simons theory, restricted on $F$ line, we have coincidence with refined topological string theory with one K\"ahler parameter,  relation (\ref{ttg}) between string coupling, K\"ahler parameter and refinement parameter, and spin content (\ref{Flineident}).

Similar result can be established for the exceptional line. We rewrite partition function as

 \begin{eqnarray}
\ln Z_{Exc} \cong \sum_{n=1}^{\infty} \frac{e^{-2\frac{g_s n}{2z}}(e^{-\frac{g_s n}{2z}}+e^{\frac{g_s n}{2z}})-e^{-11\frac{g_s n}{2z}}(e^{2\frac{g_s n}{2z}}+1+e^{-2\frac{g_s n}{2z}}) } {n (e^{\frac{g_s n}{2}}-e^{-\frac{g_s n}{2}})  (e^{\frac{g_s b n}{2}}-e^{-\frac{g_s b n}{2}})} \\
+\sum_{n=1}^{\infty} \frac{e^{-11\frac{g_s n}{2z}}} { n (e^{\frac{g_s n}{2}}-e^{-\frac{g_s n}{2}})  ( e^{\frac{g_s b n}{2}}-e^{-\frac{g_s b n}{2}})}
\end{eqnarray}

 Identify parameters in the following way, exactly the same as for F-line:
 
 \begin{eqnarray} \label{ident}
 q=e^{g_s},\,\,\,t=e^{g_s b} \\
 T_C\equiv \tau = \frac{g_s(1-b) }{2}
 \end{eqnarray}
 
 and choose 
 
 \begin{eqnarray} \label{NExc}
 N^{2B}_{0,\frac{1}{2}}=1, \,\,\, N^{11B}_{0,1}=1
 \end{eqnarray}
 
 Then Chern-Simons partition function for exceptional gauge groups (in fact, for the entire $Exc$ line) coincides with the partition function of the refined topological string with these parameters. Moreover, Chern-Simons partition function contains additional term, which we assume is a constant maps term. It differs from similar term for F-line but both coincide with the constant maps term for SU(N) case in the unrefined limit $b \rightarrow 1$ under identification (\ref{ident}), which leads to Euler characteristic $\chi(X)=2$. 
 
 This is a non-trivial coincidence. Note, for instance, the relative minus sign between spin one-half and spin one terms, which allows numbers $N^C_{j_L,j_R}$ in (\ref{Flineident}), (\ref{NExc}) to be positive. 
 
 \section{Conclusion}
 
It is interesting to consider Chern-Simons theory with the $SO(8)$ and $SU(3)$ gauge groups. Their points on Vogel's plane both belong to $Exc$ line, so they are also assumed to be dual to refined topological strings with parameters $z=-1,-2$ (i.e. $b=2, 3/2$), respectively. However, both already have their duals among unrefined topological string theories - the one on conifold, in $SU(3)$ case, and on the orientifold of conifold in the case of $SO(8)$. In other words, one can present partition function of Chern-Simons with these gauge groups in the form appropriate for forementioned non-refined topological strings. Of course this is possible only due to the restrictions of our refined topological string, namely relation (\ref{ttg}) between K\"ahler and refinement parameters, and string coupling.  Therefore, there is a possibility that can find also duals for Chern-Simons theory with exceptional groups among  non-refined, orientable or nonorientable, topological strings.

 In this study we suggest interpretation of a partition function of Chern-Simons theory on exceptional $Exc$, and $F$, lines on Vogel's plane as a partition function of the refined topological string with special parameters. It is feasible since the partition function satisfies the general structure (\ref{ztop}) for refined topological strings. This is a first step towards the gauge/string duality for Chern-Simons theories with the exceptional gauge groups. To claim  the duality one would need  to present Calabi-Yau manifold with corresponding parameters. In particular, that manifold has to have  Gopakumar-Vafa invariants $n^d_g$ calculated from above numbers $N^C_{j_L,j_R}$, e.g. for $Exc$ line non-zero invariants are  genus-zero $|n^2_0|=2, |n^{11}_0|=3$.

\section{Acknowledgments.}

I'm indebted to R. Poghossian for discussion of present paper. 

Work is partially supported by  the Science Committee of the Ministry of Science and Education of the Republic of Armenia under contract  18T-1C229.

\end{document}